# Probing Embryonic Tissue Mechanics with Laser Hole-Drilling

Xiaoyan Ma[1*], Holley E Lynch[1*], Peter C Scully[1] and M Shane Hutson[1-3]
[1]Vanderbilt Institute for Integrative Biosystem Research & Education, [2]Department of Physics & Astronomy, [3]Department of Biological Sciences, Vanderbilt University, Nashville, TN 37235

Email: shane.hutson@vanderbilt.edu

**Abstract.** We use laser hole-drilling to assess the mechanics of an embryonic epithelium during development – *in vivo* and with subcellular resolution. We ablate a subcellular cylindrical hole clean through the epithelium, and track the subsequent recoil of adjacent cells (on ms time scales). We investigate dorsal closure in the fruit fly with emphasis on apical constriction of amnioserosa cells. The mechanical behavior of this epithelium falls between that of a continuous sheet and a 2D cellular foam (a network of tensile interfaces). Tensile stress is carried both by cell-cell interfaces and by the cells' apical actin networks. Our results show that stress is slightly concentrated along interfaces (1.6-fold), but only in early closure. Furthermore, closure is marked by a decrease in the recoil power-law exponent – implying a transition to a more solid-like tissue. We use the site- and stage-dependence of the recoil kinetics to constrain how the cellular mechanics change during closure. We apply these results to test extant computational models.

## 1. Introduction

Morphogenesis in developmental biology is clearly both a genetic and a mechanical process. The mechanical aspects are the proximate cause of the main observables – shape and form – but typically receive scant attention. When developmental mechanics are considered, the normal course is for physical scientists to develop computational models [1-6]. These models reproduce the shapes and forms of morphogenesis with varying degrees of success and complexity. Ideally, such models would be challenged and refined with complementary experiments. Unfortunately, the experimental tools of developmental biology are ill equipped to test mechanical hypotheses. Such hypotheses can and have been tested using laser microsurgery [7-12]. Here, we show that a variant of microsurgery, referred to as laser hole-drilling, can elucidate how the subcellular stress distribution and mechanical properties change during a specific episode of morphogenesis – the apical constriction of amnioserosa cells during dorsal closure.

At the beginning of dorsal closure (Bowne's stage 13 in *Drosophila* [13]), the embryo's dorsal surface is covered by a one-cell thick epithelium, the amnioserosa. As closure proceeds (stages 13-15), the adjacent epidermis advances from the embryo's lateral flanks to seal over the amnioserosa, which dives inside the embryo. Dorsal closure has been extensively characterized in terms of its genetics, cell shape changes and tissue-level mechanics [7, 14-17]. Laser microsurgery played a large role in the latter and established critical roles for three coordinated processes [7, 16]: an adhesive interaction between approaching flanks of epidermis; contraction of a supracellular "purse-string" along the amnioserosa-epidermis boundary; and apical constriction of amnioserosa cells. Similar instances of apical constriction occur in gastrulation and neurulation [18, 19]. In each case, a subgroup of epithelial cells contracts their apical surfaces while expanding in the apical-basal direction. When viewed in cross-section, the constricting cells take on a wedge-shaped morphology. Such constriction contributes most of the force for dorsal closure [7], and is the subject of our analysis here.

---
[*] These authors contributed equally to this work.

Our goal is to test, validate and eliminate alternative models. In general, epithelial sheets have been modeled as 2D cellular foams – i.e. a network of tensile cell-cell interfaces [4, 10, 20]. The cytoplasm, apical and basal surfaces are treated as passive, incompressible media. The first major question to address is whether the amnioserosa behaves more like a foam or a continuous sheet. Beyond this general consideration, apical constriction has been reproduced by multiple models including: stretch-induced apical contraction [1]; programmed, time-dependent changes in interfacial tensions [2] or in tissue-level tensile stresses [3]; compressive forces applied by cells outside the constricting region [3]; and a combination of programmed changes in tensions and the physics of embryonic hydrodynamics [6]. Each model drives constriction through a different set of time-dependent stresses and mechanical properties. The experiments below provide constraints on how the stresses and mechanical properties actually change during one example of apical constriction. In the conclusions, we evaluate the models in light of the experimental constraints.

The experiments presented here are inspired by hole-drilling methods for evaluating residual stress [21, 22]; however, those methods estimate stress from equilibrium elastic deformation. For a viscoelastic material like a cell sheet, the primary information is in the recoil dynamics. Thus, for all of our experiments, hole-drilling consists of ablation with a single laser pulse (at 355 nm, 2-3× threshold). Under these conditions, the longest-lived ablation transients are due to cavitation. The bubbles extend out < 5 µm and collapse in < 2 µs [23]. Most of the bubble expansion is into the extracellular space, leaving a damaged cellular region of just a few µm. Such localized, impulsive ablation is critical for measuring the complete recoil dynamics. These dynamics are the key to evaluating how the stresses and mechanical properties vary within an epithelium and between developmental stages.

## 2. Materials and Methods

### 2.1 Fly strains and microscopy.
The primary strain of *Drosophila melanogaster* used in this study is ubi-DE-Cad-GFP [24] (Kyoto Drosophila Genetic Resource Center). This strain ubiquitously expresses a cadherin-GFP chimera that labels epithelial cell junctions. A few experiments use the strain sGMCA [16] (gift from D.P. Kiehart), which expresses a GFP-moesin construct that labels filamentous actin. For imaging and ablation, fly embryos were dechorionated in 50% bleach solution, immersed in halocarbon oil 27 (Sigma-Aldrich, St. Louis, MO), and sandwiched between a cover glass and an oxygen-permeable membrane (YSI, Yellow Spring, OH) [25]. Images were captured on a Zeiss LSM410 laser-scanning confocal microscope (inverted) with a 40×, 1.3 NA oil-immersion objective and 488-nm excitation. The scanning times were 2-8 s per frame and 15.74 ms per kymograph line.

### 2.2 Laser microsurgery.
Ablations were performed with the 3$^{rd}$ harmonic (355 nm) of a Q-switched Nd: YAG laser (5-ns pulsewidth, Continuum Minilite II, Santa Clara, CA). This laser was coupled into the Zeiss LSM410 with independent beam steering for simultaneous ablation and imaging [25]. The pulse energy was just high enough (2-3× threshold) to insure consistent single-pulse ablation. This energy varied with embryo stage and tissue – 1.16 ± 0.25 µJ for amnioserosa cells in stage 13 and 3.33 ± 0.91 µJ for these cells in stage 14 – due to differences in depth below the embryo's vitelline membrane – 2.3 ± 1.7 versus 16.4 ± 6.1 µm, respectively.

**2.3 Image and data analysis.**

All image processing was performed with ImageJ (NIH, Bethesda, MD). Specific plugins were used to extract the spatial displacement patterns from before-and-after images (UnwarpJ [26]) and the time dependence from kymographs (custom, based on the Lucas-Kanade algorithm [27]).

The recoil kinetics were parameterized via non-linear regression. We did not independently measure the time at which ablation occurred, $t_0$, so this time was estimated by fitting each data set to a piecewise continuous function

$$x(t) = d_0 + m(t - t_0) \qquad\qquad\qquad\qquad\qquad\qquad t < t_0$$
$$x(t) = d_0 + d_1\left(1 - e^{-(t-t_0)/\tau_1}\right) + d_2\left(1 - e^{-(t-t_0)/\tau_2}\right) \qquad t \geq t_0$$

Eqn 1.

This functional form gave the most reliable estimates of $t_0$ for a variety of simulated data. Once we obtained fitted estimates for $d_0$ and $t_0$, the displacements for $t \geq t_0$ were then fit to a power-law,

$$\Delta x(t) = x - d_0 = D(t - t_0)^\alpha \qquad\qquad \text{Eqn 2}$$

where only $D$ and $\alpha$ were adjustable. For presentation, all graphs have been shifted so that $t_0 = 0$.

For further analysis, the data was grouped by ablation site (cell center versus cell edge) and developmental stage (13 versus 14). For each group, the mean parameter values are reported as mean ± standard error of the mean (using the standard errors from each experiment and the variance among experiments). To assess the significance of site and stage differences, we used a two-factor ANOVA. To assess any dependence on the tissue-level alignment of cell edges, we used a two-factor ANCOVA with a single covariate (angular distribution of cell edges) [28]. To insure statistical independence among the samples, each analysis considered just one randomly selected cell edge from each embryo. Regression and statistical analysis was performed in Mathematica (Wolfram Research, Champaign, IL).

To display the parameter distributions for each group, we constructed kernel density estimates using a variable-width Gaussian kernel [29, 30]. Kernel density estimates are similar to a histogram, but avoid the artifacts associated with choosing specific bin boundaries. The kernel width, $h$, is analogous to a histogram bin width and was chosen to slightly undersmooth the density estimates (as noted on each graph). In a few experiments, the standard error of a parameter exceeded $h$. In such cases, the kernel width was broadened to equal the standard error. A similar kernel density estimate was used to estimate the density distribution of cell edge orientations (using a fixed $h = 7°$). Since this distribution is with respect to a periodic angular variable, we wrapped the distribution using one period above and below that shown [31].

## 3. Results and Discussion

Before addressing questions of epithelial mechanics, we need to establish the extent of damage caused by hole-drilling. Fig 1 shows two examples. The initial hole in the epithelium is marked by a similar hole in the embryo's overlying vitelline membrane (e.g. the dark region with a hyper-fluorescent ring in Fig 1E). This hole in the vitelline is bounded a glue layer that holds the embryo onto the coverslip. Thus, it expands little and does not allow material to flow in or out. The hole apparent in Fig 1E is elliptical with semiminor and semimajor axes of 0.50 and 0.75 µm – approximately one-tenth the size of a typical cell. In contrast, the hole in the epithelium rapidly expands. The wounded cell(s) maximally expand by about ½ a cell diameter in 20-60 s (average velocity ~ 1 µm/s). The neighboring cells then move back towards the ablation site as wound healing begins.

In cross-section, the laser wounds cut cleanly through the epithelium (~ 6 µm thick), allowing the apical and basal surfaces to both move freely. There is some apparent shear in the first cross-section image after wounding (Fig 1G), but this is just a motion artifact. The apparent shear is reversed when the cross-sections are collected from bottom-to-top (Fig 1H). Thus, our laser protocol creates a hole clean through the epithelium that is 5-7 µm deep.

To then address whether the amnioserosa behaves more like a 2D foam or a continuous sheet, we performed single-pulse laser ablation of the stage-13 amnioserosa at various subcellular locations. Most importantly, strong recoils occur regardless of whether the wound is targeted to a cell edge (Fig 1A-C) or cell center (Fig 1D-F). Apparently – and in contrast to common assumptions [4, 16, 32] – tension is not limited to the actin-rich apical belts around each cell.

To further investigate the cell-center wounds, we moved the ablation focus perpendicular to the plane of the epithelium. Strong recoils occurred whenever the laser cut a cell's apical surface – regardless of whether it also cut the basal surface ($N = 16$, supplemental Table S1). Recoils did not occur when the laser cut just the vitelline membrane. Thus, cell-center tension is carried by structures within a few µm of the apical surface. These structures are not specifically associated with the nucleus because strong recoils occurred regardless of whether the nucleus was targeted or not ($N = 5$ and 8, respectively). We will return to identifying the tension-carrying structures in Section 3.8.

To help interpret the recoils, we also performed a series of double-wounding experiments. The second ablation occurs as the initial recoil slows to a halt. If the second ablation targets the same location, there is no recoil ($N = 2$, supplemental Movie S3); however, if it targets an adjacent cell or even a different location in the same cell(s), there is always a second recoil ($N = 13$ and 16 respectively, Movies S4 and S5). The wound does not stop expanding because tension relaxes across the epithelium. Quite to the contrary, wound expansion slows and eventually stops as the local stress decreases in the radial direction, but increases in the azimuthal direction. Furthermore, a hole is not equivalent to releasing tension across an entire ablated cell(s).

**3.1 Spatial dependence of the relaxed displacements.**
Since the amnioserosa does not behave like a 2D foam, we next compare the recoil patterns to those expected for a continuous sheet. We track the spatial recoil pattern by measuring the displacements of specific cellular triple junctions (bold arrows in Fig 2) and by performing an elastic sheet registration of pre- and post-ablation images (lighter arrows). The elastic sheet registration uses B-splines with regularization conditions to approximate any smooth elastic deformation [26]. This functional approximation also enables differentiation to find the relaxed strain. With either measure, the recoil pattern is generally radial, but has a weak anisotropy that follows the local arrangement of cell edges. The influence of adjacent cell edges is clearest in the radial plots of relaxed strain. The directions with the most negative relaxed strain coincide with the ablated edge in Fig 2B and with the adjacent cell edges in Fig 2D. The weakness of this anisotropy is evident in Fig 2A, where the triple junctions at the ends of the wounded edge do recoil farther than any other triple junctions of the same cell, but only by 5%.

For comparison, consider the analytic solution for a circular hole through a thin sheet under biaxial plane stress ($\sigma_x$, $\sigma_y$). If one assumes a homogeneous sheet that is isotropic, linearly elastic, and undergoes infinitesimal deformations, then the relaxed strains are given by [33, 34]

$$\Delta\varepsilon_r(r,\theta) = -A_1(r)(\sigma_x + \sigma_y) - A_2(r)(\sigma_x - \sigma_y)\cos 2\theta$$
$$\Delta\varepsilon_\theta(r,\theta) = A_1(r)(\sigma_x + \sigma_y) + A_3(r)(\sigma_x - \sigma_y)\cos 2\theta$$
Eqn 3

in polar coordinates centered on the hole with $\theta$ measured from the direction of principal stress. The dependence of relaxed strain on distance is contained in the coefficients (for $r \geq R_0$):

$$A_1(r) = \frac{1+\nu}{2E}\left(\frac{R_0}{r}\right)^2 \qquad A_2(r) = \frac{1+\nu}{2E}\left[\frac{4}{1+\nu}\left(\frac{R_0}{r}\right)^2 - 3\left(\frac{R_0}{r}\right)^4\right] \qquad A_3(r) = \frac{1+\nu}{2E}\left[\frac{4\nu}{1+\nu}\left(\frac{R_0}{r}\right)^2 - 3\left(\frac{R_0}{r}\right)^4\right] \quad \text{Eqn 4}$$

where $R_0$ is the radius of the hole, $\nu$ is Poisson's ratio and $E$ is Young's modulus. This special case underlies the standard engineering method for determining residual stress [22]. In the isotropic case, expansion of the hole leads to a decrease in radial stress and stain ($\Delta\varepsilon_r < 0$) and an increase in azimuthal stress and strain ($\Delta\varepsilon_\theta > 0$).

The corresponding displacements include a possible rigid body translation (of magnitude $u_{tr}$ and direction $\theta_{tr}$ caused by asymmetry in the boundary conditions on the sheet):

$$u_r(r,\theta) = B_1(r)(\sigma_x + \sigma_y) + B_2(r)(\sigma_x - \sigma_y)\cos 2\theta + u_{tr}\cos(\theta - \theta_{tr})$$
$$u_\theta(r,\theta) = -B_3(r)(\sigma_x - \sigma_y)\sin 2\theta + u_{tr}\sin(\theta - \theta_{tr})$$
Eqn 5

where

$$B_1(r) = \frac{1+\nu}{2E}\frac{R_0^2}{r} \qquad B_2(r) = \frac{1+\nu}{2E}\left[\frac{4}{1+\nu}\frac{R_0^2}{r} - \frac{R_0^4}{r^3}\right] \qquad B_3(r) = \frac{1+\nu}{2E}\left[2\frac{1-\nu}{1+\nu}\frac{R_0^2}{r} + \frac{R_0^4}{r^3}\right]$$
Eqn 6.

Although our experiments violate the idealized assumptions – most notably because the cell sheets are viscoelastic and undergo finite deformations – this model provides a useful framework. We also note that cells are an inherently "active" material. The above relationships would only hold during the cells' initial passive response phase; and the switch from passive to active response could happen on time scales < 1 s [35]. Nonetheless, when we compare the distance dependence, the experimental displacements do fall off as approximately $1/r$. When we compare the angular dependence, the experimental displacements display both a rigid body shift and a $\cos 2\theta$ anisotropy.

To be more specific, we measured $u_r(r,\theta)$ for each triple-junction visible in Fig 2 and fit these displacements to Eq 5. The best fits both had rigid body shifts to the right: 0.79 ± 0.07 μm at 7 ± 6° for Fig 2A; and 0.83 ± 0.12 μm at 22 ± 9° for Fig 2D. A larger sampling of experiments shows that rigid shifts are usually along the embryo's anterior-posterior axis. The size and direction should be determined by where the embryo contacts the overlying and rigidly held vitelline membrane. When the rigid body shifts are removed by calculating relaxed strains, the polar plots resemble ellipses (Fig 2B,E), as expected for a $\cos 2\theta$-dependence. In the idealized case, the long axis of each ellipse would be parallel to the far-field principal stress; however, these axes also coincide with local features, e.g. the ablated edge or long axis of the ablated cell. Furthermore, for $r < 15$ μm, the elliptical shape breaks down. In Fig 2E, the relaxed strain actually follows a tri-lobed pattern that coincides with the adjacent cell edges. Beyond the first ring of cells, the pattern is nearly isotropic. This isotropy appears in the best fits to Eq 5, where the coefficients of the $\cos 2\theta$ terms are less than 3% of the leading isotropic terms.

To assess the distance dependence of $u_r$, we used a simpler, isotropic version of Eq 5 ($\sigma_x = \sigma_y = \sigma$). This leaves just three fitting parameters ($u_{tr}$, $\theta_{tr}$ and a combination of factors $C = (1+\nu)R_0^2\sigma/E$ in the coefficient to $1/r$), but still fits the data well (Fig 2C,F). For the isotropic case, the coefficient $C$ is related to the pre-ablation strain in the cell sheet as

$$\varepsilon_{r,0} = \frac{\sigma}{E}(1-\nu) = \frac{C}{R_0^2}\frac{1-\nu}{1+\nu}$$
Eqn 7.

If one assumes incompressibility ($\nu = \frac{1}{2}$) and uses the size of the hole in the vitelline ($R_0 = 0.5$ μm), then the estimated pre-ablation strains are unreasonably large ($\varepsilon_{r,0} > 40$). The hole in the cell sheet must be larger than that in the vitelline, but it must also be smaller than the full volume of the ablated cell(s). As discussed below, double-wounding experiments show that the first hole in a cell does not release all of its tension. In fact, experiments on GFP-moesin embryos allow one to see multiple subcellular holes in a single cell's apical actin network. From these images, we estimate an upper limit on cellular hole size ($R_0 < 5$ µm) that implies a more reasonable lower limit on the pre-ablation strains ($\varepsilon_{r,0} > 0.4$ and 0.7 for Fig 2A,D respectively).

Overall, the spatial pattern falls between the two idealized cases. Beyond the first ring of cells, the displacements match a homogeneous thin sheet model. At closer distances, the displacements are influenced by the local arrangement of cell edges. To further quantify mechanical inhomogeneity at the cell level, we next turn to the recoil kinetics.

**3.2 Temporal dependence of the relaxed displacements.**
By 20 s after ablation, the recoils have slowed dramatically, but they are not in a true static equilibrium. In fact, expansion of the ablated hole only pauses transiently before wound healing commences. Thus, recoil patterns on the 10-s time scale may be contaminated by secondary effects, i.e. biologically regulated changes in the cytoskeleton [35]. More direct information with regard to the epithelium's viscoelastic properties and its local stresses is contained in the short-time recoil kinetics.

Our full-frame confocal images are too slow to measure these kinetics (>2 s per frame), so we limited our measurements to repeated scans along a single line (one scan every 15.7 ms). These line scans are used to construct a recoil kymograph (Fig 3B). Before ablation, the kymograph has a series of bright vertical bands – one for each nearly immobile cell edge. Immediately after ablation, the bands fan out as the cells recoil away from the ablation site.

For the cell edges closest to a wound (not counting the wounded edge itself), the recoil displacements show multiphasic behavior (Fig 3CD). Over two decades (0.1 to 10 s), the recoils are well fit by a weak power law (exponent 0.2-0.6). At shorter times, the recoils fall slightly below this power law, consistent with a transient linear regime. Despite the problems at very short times, a power law was able to fit each measured recoil ($N$=179) with an adjusted $R^2$ greater than 0.966. We could achieve equivalent $R^2$ and a better fit at short times using a double exponential; however, its two fitted time constants depend strongly on the time range used for the fits: 40-80 ms and 1-2 s when fitting two seconds of data, but 200-400 ms and 5-10 s when fitting ten seconds. The two power law parameters provide a more succinct and consistent description. We will take a closer look at the fit parameters below, but will first make some model-independent recoil comparisons.

These comparisons are designed to assess both the sub-cellular distribution of stress and how this stress changes during development. We measured recoil kymographs for both cell-center and cell-edge wounds in stage 13 and 14 embryos (early and late dorsal closure). We made one wound per embryo and tracked the closest cell edges on both sides of the wound (if possible). The numbers of embryos wounded and edges tracked are listed in Table I. Fig 4 then compares the average recoils for each stage/site combination. In stage 13, the cell-edge wounds recoil more than cell-center wounds (~1.6×). In stage 14, they do not. Interestingly, despite the differences in recoil extent, the slopes on a log-log plot (Fig 4B) are similar for both wound sites. As closure progresses, the major stage-dependent changes are a decrease in the log-log slope and a large reduction in the extent of cell-edge recoils.

To quantify this apparent stage and site dependence, we fit a power law to each experimental recoil. For each stage/site combination, the mean parameters are reported in Table I. Using a two-factor ANOVA, the power law prefactor $D$ is clearly stage-dependent, site-dependent and stage/site-codependent (all $P < 0.01$). The mean values of $D$ reflect the differences in recoil extent noted above – 1.6× larger for cell-edge wounds in stage 13, but roughly equivalent in stage 14. On the other hand, two-factor ANOVA shows that the power law exponent $\alpha$ is only dependent on developmental stage ($P = 1\times10^{-4}$), not on wound site ($P = 0.47$). As closure progresses, $\langle\alpha\rangle$ decreases from 0.39 to 0.32. These dependences are reflected in the kernel density estimates of the parameter distributions (Fig 5AB, corresponding histograms in supplemental Fig S1). For $D$, the distribution of stage 13 cell-edge wounds clearly stands apart. For $\alpha$, the wound-site distributions strongly overlap in a single stage and coordinately shift between stages.

| Category | $N(M)$ | $D$ (μm/s$^\alpha$) | $\alpha$ | $v_0$ (μm/s) |
|---|---|---|---|---|
| Stage 13: center | 51 (30) | 1.34 ± 0.07 | 0.396 ± 0.015 | 13.4 ± 1.5 |
| edge | 58 (36) | 2.17 ± 0.13 | 0.381 ± 0.010 | 20.0 ± 2.0 |
| Stage 14: center | 33 (22) | 1.52 ± 0.12 | 0.305 ± 0.013 | 24.7 ± 2.6 |
| edge | 37 (25) | 1.36 ± 0.07 | 0.329 ± 0.018 | 19.4 ± 2.1 |
| ANOVA/ANCOVA | | | | |
| Stage | | 0.01 | $1 \times 10^{-4}$ | 0.02 |
| Site | | $6 \times 10^{-3}$ | – | – |
| Stage/Site | | $8 \times 10^{-4}$ | – | 0.02 |
| $\rho(\theta)$ | | – | – | $1 \times 10^{-3}$ |

**TABLE I.** Summary of recoil parameters. The top half of the table reports each parameter as its mean ± standard error of the mean. For each category, $N$ is the number of edges tracked and $M$ is the number of embryos wounded. The bottom half of the table reports the $P$-values from ANOVA ($D$ and $\alpha$) or ANCOVA ($v_0$) for any significant factor or covariate dependence ($P < 0.05$).

### 3.3 Initial recoil velocities.

Previous microsurgery experiments focused on the initial recoil velocity $v_0$ as a probe of cell and tissue mechanics [7-10]. Unfortunately, the power law fits cannot be used to estimate $v_0$ because the limiting slope is infinite. Instead, we assessed $v_0$ by fitting the recoils to a double exponential (Eq 1). In contrast to the power law, this function fits the early phase of recoil quite well (Fig 3D). The mean $v_0$ for each stage/site combination is listed in Table I. A two-factor ANOVA shows that $v_0$ is stage/site-codependent ($P = 0.02$). In stage 13, $\langle v_0 \rangle$ is 1.5× faster for cell-edge wounds. In stage 14, $\langle v_0 \rangle$ is similar for both wound sites. This dependence is very similar to that found for the power law prefactor.

Kernel density estimates for the $v_0$-distributions are shown in Fig 5D. These distributions do not appear to be singly peaked. To explain the multiple peaks, we looked for correlations between $v_0$ and several geometric factors. Most gave negative results. For example, we found no correlation between $v_0$ and any of the following: length of the ablated edge; area of the ablated cell(s); perimeter of the ablated cell(s); and angles at the triple-junctions adjacent to an ablated edge.

The only apparent correlation is between $v_0$ and the tissue-level alignment of cell edges, as measured by the density of cell edges in each direction, $\rho(\theta)$. As shown in Fig 6A, the overall pattern of cells changes drastically during dorsal closure. In stage 13, the cells are roughly isodiametric and hexagonal. By stage 14, the cells are longer in the AP direction and diamond-shaped. We measured the orientations of 229 cell edges in stage 13 and 157 edges in stage 14 to construct kernel density estimates for $\rho(\theta)$, shown in Fig 6B. The quasi-hexagonal cells in stage-13 have an angular density distribution with three nearly equal peaks near 30, 90 and 150°. The diamond-shaped cells in stage-14 have a drastically reduced peak near 90° and much larger peaks near 20 and 160°. When $v_0$ is plotted against the direction in which recoil was tracked (Fig 6CD), the largest velocities line up with the peaks in the stage-dependent $\rho(\theta)$-distribution. Note that we tracked the recoil in only one direction for any one ablation. The angular pattern emerges only after tracking different recoil directions for many different wounds.

To assess the significance of this apparent correlation between $v_0$ and the density of parallel cell edges, we performed a two-factor, one-covariate ANCOVA. Note that we are not testing the obviously non-linear relationship between $v_0$ and $\theta$ (Fig 6CD), but a linear relationship between $v_0(\theta)$ and $\rho(\theta)$. As Fig 6CD makes clear, the angular-dependence of $v_0$ changes with developmental stage, but we want to test the hypothesis that this change is due to the tissue-level realignment of cell edges. To perform a similar ANOVA with categorical factors, the factor levels would need to be the stage-dependent and non-contiguous angular ranges that correspond to high, medium and low cell edge density. The ANCOVA results are listed in Table I and confirm the correlation between $v_0$ and $\rho(\theta)$ at $P = 1\times10^{-3}$ (with a best-fit slope of $1950 \pm 570$ µm°/s). Beyond this relationship, the analysis also confirms the stage/site-codependence of $v_0$. We performed similar ANCOVA for the power law parameters, but found no significant correlation between $\rho(\theta)$ and $D$ or $\alpha$.

Note that we have used impulsive ablation (a single 5-ns pulse) and fast time resolution ($\Delta t = $ 15.7 ms) to measure initial recoil velocities of 5-50 µm/s. Such velocities are 10-100× faster than those reported in other microsurgery experiments: 1-3 µm/s for multi-cell ablations in the amnioserosa [7-9]; < 0.3 µm/s for ablation of single cell edges in the fly wing disk [11] or extending germ band [10]; and 0.5-1 µm/s for ablation of single stress fibers in cultured cells [12]. Part of the discrepancy is due to our faster time resolution. Previous experiments estimated $v_0$ over a time window of several seconds; and if we calculate the average recoil velocity during the first two-seconds, $\langle v_{0-2s} \rangle$, we too find speeds of ~1 µm/s. We would thus expect higher time resolution to lead to higher revised estimates of $v_0$ in previously examined tissues. Even with a 10× upward revision, there would still be a 5-10× discrepancy with our results that is likely due to tissue and developmental stage effects. In preliminary experiments during fly stages 11-12 (germband retraction), we observe recoil velocities that are an order of magnitude smaller than those reported here. The amnioserosa during dorsal closure appears to be under much more tensile stress than other tissues.

Although $\langle v_{0-2s} \rangle$ is much lower than the true $v_0$, its stage and site comparisons are similar to those for $v_0$ and $D$ (Fig 5C) – faster for cell-edge wounds in stage 13 ($1.41 \pm 0.09$ vs $0.88 \pm 0.06$ µm/s), but with little difference between wound sites in stage 14 ($0.91 \pm 0.05$ versus $0.95 \pm 0.08$ µm/s). In contrast to $v_0$, and more like $D$, ANCOVA shows that $\langle v_{0-2s} \rangle$ has no particular dependence on $\rho(\theta)$ (supplemental Fig S2). The faster recoils thus provide some information that is unavailable on longer time scales.

**3.4 Mechanical interpretation of the recoil kinetics.**
What do the above results tell us with regard to the cell-level and stage-dependent mechanics of the amnioserosa? Recent microrheometry has recognized that cells behave like soft, glassy materials – with a continuous spectrum of relaxation times and a creep function with power-law behavior [36-40]. For materials that follow power-law rheology, the sudden application of a constant stress $\sigma$ leads to a subsequent time-dependent strain (assuming a linear response):

$$\varepsilon(t) = \frac{\sigma}{G_0}\left(\frac{t}{\tau_0}\right)^\alpha \qquad \text{Eqn 8}$$

where $G_0$ and $\tau_0$ are scale factors for stiffness and time respectively [39]. The exponent $\alpha$ varies between 0 and 1 (purely elastic solid to purely viscous fluid), with smaller exponents implying more solid-like materials [40].

Our recoils are similar to a creep experiment with some strong caveats. First, instead of applying a constant stress, we suddenly remove a local stress – equivalent to the sudden application of an extra local stress of opposite sign. Second, the geometry of our experiments (radial expansion) are very different from most others (uniaxial stress or a local torque). Third, the analogy breaks down when the recoil displacements cause a relaxation in the far-field stress. Such relaxation is unavoidable in a 1D experiment like nanodissection of stress fibers [12], but can be quite small in two dimensions. As shown in the double-wounding experiments, expansion of the hole slows down due to increases in the local azimuthal stress, not decreases in the far-field stress. Even so, the analogy breaks down when the cells begin to actively remodel during wound healing (20-60 s after ablation).

To compare our results to Eq 8, we use the measured displacements to approximate the relaxed strains. Since the displacements fall off as $1/r$, the relaxed strains go as $-u/r$. We thus approximate $-D/r_0 \approx \Delta = \sigma/(G_0\tau_0^\alpha)$ where $r_0$ is the distance from the ablation site to the tracked edge. This approximation makes almost no difference for intra-stage comparisons, but impacts inter-stage comparisons because $\langle r_0 \rangle$ gets smaller in late dorsal closure (Table II).

At early times after ablation, the observed recoils do not follow power law behavior, but are instead nearly linear. This is consistent with a short time regime that is dominated by a Newtonian viscosity. Thus, we can also relate $v_0$ to the locally removed stress by $-v_0/r_0 \approx \gamma_0 = \sigma/\eta$ where $\gamma_0$ is the initial strain rate and $\eta$ is a viscous drag coefficient.

The recoil strain parameters are thus determined by both the pre-ablation stress at the hole and the post-ablation properties of the surrounding tissue. Since the hole is a very small fraction of the entire epithelium, $\sim 10^{-4}$, the mechanical properties of the remaining tissue should be nearly independent of the hole's location. This supposition is supported by the lack of a site-dependence for $\alpha$. Within a single stage then, $\Delta$ and $\gamma_0$ directly report on the cellular stress distribution. Both suggest an uneven distribution in early dorsal closure (~1.6× more stress along cell edges) that evolves into a more uniform loading.

Our present results cannot uniquely define how this evolution occurs. Cell-edge stress could decrease, or cell-center stress could increase. One can capture the coupled constraints through a grand ratio of the stage-dependent changes in both stresses: $R_{\sigma C}/R_{\sigma E} = (\sigma_{C14}/\sigma_{C13})/(\sigma_{E14}/\sigma_{E13}) = 2.06 \pm 0.28$. The changes in mechanical properties are similarly constrained as noted in Table II (details in Supplemental Note 1). As just one example of a consistent scenario, consider a case where the viscous drag coefficient does not change. The constraints then require increases in the amnioserosa stiffness (1.6×), cell-edge stress (1.3×) and cell-center stress (2.7×). We cannot tighten the constraints to a single scenario without independent estimates of one or more mechanical properties.

| Stage/site | $r_0$ | $\Delta$ (s$^{-\alpha}$) | $\gamma_0$ (s$^{-1}$) | Relative Changes | |
|---|---|---|---|---|---|
| | | | | Stage 14 vs 13 | |
| 13/center | 7.55 ± 1.60 | 0.189 ± 0.014 | 1.84 ± 0.21 | | |
| 13/edge | 7.64 ± 1.59 | 0.295 ± 0.018 | 2.74 ± 0.29 | $R_{\sigma C}/R_{\sigma E}$ | = 2.06 ± 0.28 |
| 14/center | 5.24 ± 1.35 | 0.307 ± 0.027 | 5.04 ± 0.55 | $R_{G'}/R_{\sigma E}$ | = 1.24 ± 0.07 |
| 14/edge | 5.82 ± 1.35 | 0.242 ± 0.013 | 3.49 ± 0.36 | $R_\eta/R_{\sigma E}$ | = 0.77 ± 0.08 |

**TABLE II.** Estimated relaxed strain parameters and grand ratios summarizing how the stresses and mechanical properties coordinately change during dorsal closure.

The one unambiguous, stage-dependent change in the mechanics is a decrease in the power-law exponent. As closure progresses, $\alpha$ decreases as the amnioserosa becomes more solid-like. The above constraints are consistent with the change in $\alpha$; the stiffness always increases more (or decreases less) than the viscous drag. To our knowledge, this is the first report in which cells have been shown to regulate this exponent during a morphological change *in vivo*. The exponents we measure are a bit larger than those reported from microrheology and single-cell stretching rheology of cultured cells, 0.1-0.3 [38-41], but embryonic epithelia have a less-developed extracellular matrix and more fluid-like behavior [42].

**3.8 The apical actin network in amnioserosa cells.**
Certainly, the most surprising finding here is the subcellular stress distribution. Like many epithelia, amnioserosa cells have their actin cytoskeletons organized cortically – largely in circumferential microfilament bundles. This organization is ubiquitously assumed to yield concentrated tensile forces along cell-cell interfaces. Whether this assumption is stated explicitly [32] or not [16], it is often used to estimate relative tensions based on the angles at cell triple-junctions. Despite widespread use of this assumption, even in other laser ablation studies [10], our results suggest that it is not applicable to all epithelia. The amnioserosa does not behave like a 2D foam, but more like a continuous sheet.

In addition to the circumferential microfilaments, amnioserosa cells have an apical actin network. Others have reported transient apical accumulations of actin and myosin in amnioserosa cells, especially those that dive out of the epithelial plane early [9, 16, 43]. Similar accumulations are evident in Fig 7A, which shows confocal images from basal, middle and apical planes of the amnioserosa in a GFP-moesin embryo. In the middle plane, one can see the actin-rich belts along cell edges. In the apical plane, the cell centers are more fluorescent and the cell borders are covered with a carpet of very mobile, actin-based projections. The basal surface also has actin-based projections, but they are not as mobile.

In time-lapse images of the apical surface (supplemental Movie S6), one can even see traveling contraction waves coupled to apical actin accumulations. This coupling is apparent in long-time kymographs of the amnioserosa (Fig 7B). The vertical lines correspond to cell edges that staircase back and forth with time. The turning points of this movement match up with bright horizontal lines – i.e. transient local increases in actin. The overall effect makes the kymograph look like a ladder. The wave speed and period are approximately 0.2 µm/s and 200-300 s.

To investigate the role of the apical network in the recoil mechanics, we performed a double-wounding experiment on a GFP-moesin embryo (Fig 7C, supplemental Movie S7). By 24 s after the first ablation, there is a clear hole in the apical actin network (dark region around the hyperfluorescent mark on the vitelline). This hole is ~10 µm across – larger than the laser spot, but smaller than a cell. Just as in Fig 1, the edges of the ablated cell have recoiled away from the wound site. At 48 s after the first ablation, we ablate a different location in the same cell. There is an immediate second recoil and formation of a second hole in the actin network. These results, combined with the correlation of actin accumulation and cellular contraction, point to the apical actin network as the carrier of cell-center stress.

**3.9 Implications for apical constriction models.**
Taking all of the results into account, the mechanics of apical constriction in dorsal closure can be summarized as follows. First, the amnioserosa more closely resembles a continuous sheet than a network of tensile cell edges. Tensile stress is carried by both circumferential actin and the apical actin network. Nonetheless, the local and global arrangement of cell edges does influence the recoil patterns. This is most evident in the strain pattern around a hole, and in the correlation between $v_0$ and the tissue-level arrangement of cell edges. Second, the recoil kinetics after ablation follow power-law behavior. Interestingly, the power-law exponent decreases as closure progresses, similar to a sol-gel transformation, and indicating solidification of the cytoskeleton. Third, closure is accompanied by a redistribution of tensile stress. An early concentration along cell-cell interfaces evolves to a more uniform distribution. The experiments here do not provide a unique description of the redistribution, but do place quantitative constraints on how the stresses and mechanical properties coordinately change.

The constraints can be applied to evaluate existing models of apical constriction. Although these models were each developed for other morphogenetic events, one can ask whether the invoked mechanisms could validly apply to dorsal closure. As a first example, consider a set of finite-element (FE) models for sea urchin invagination that evaluates several alternative mechanisms [3]. One set of models drives apical constriction of the invaginating cells by applying a compressive force from non-invaginating cells (cell tractor, apical contractile ring and gel swelling models). These mechanisms are not consistent with our observation that stress in the constricting amnioserosa cells is always tensile. This discrepancy might be alleviated if the models explicitly added an inflation pressure from the enclosed yolk. Another set of models does generate tensile stresses (via active apical constriction and apicobasal contraction), but each is modeled by positing an initial strain that drives constriction as it relaxes – i.e. a constant elastic modulus, but a stress that decreases with time. Such models are not consistent with our experiments. If the modulus $G_0$ were constant during constriction, then the cell-center stress would increase by 1.7×. As a second example, FE models of neurulation drive apical constriction by positing a locally higher interfacial tension which yields more realistic cell shapes if the tension increases with time [2]. This is possible within the constrained scenarios presented here, but would require simultaneous and even larger increases in the cell-center stresses and tissue stiffness. A similar comparison holds for models of invagination and neurulation based on stretch-induced apical contractions [1]. These models also lead to stresses that increase with time as more cells are stretched beyond a threshold and actively "fire" an apical contraction. The contractile waves evident in Fig 7 are suggestive of the amnioserosa as an active mechanical medium. Unfortunately, in all of these examples, our ability to compare experiments and models is limited by what the models explicitly report. This is usually just the cell shape changes, but it would be very useful to provide the corresponding time-dependent stresses.

## 4. Conclusions and Outlook

Although each apical constriction model produces simulated cell shapes that match observations, our laser hole-drilling results provide an even more stringent test. Given the growing utility of laser microsurgery, we hope future models of developmental mechanics will include explicit microsurgical predictions. To further constrain the models, we will need to couple laser hole-drilling with microrheology and genetics. Complementary microrheology is needed to narrow the constraints to a single mechanical scenario and provide absolute estimates of the cellular stresses. Complementary genetics is needed to integrate the molecular and mechanical aspects of development.

## Acknowledgements

This work supported by the National Science Foundation (IOB-0545679) and the Human Frontier Science Program (RGP0021/2007C). The authors thank G.W. Brodland, of U. Waterloo for many helpful discussions.

## Glossary

*amnioserosa* – a one-cell thick embryonic epithelium that covers most of the dorsal surface of *Drosophila* embryos in the latter half of embryogenesis

*apical constriction* – morphogenetic event in which a subset of epithelial cells contract their apical surfaces to adopt a wedge-shaped morphology and drive a local invagination of a tissue

*azimuthal direction* – applies to a cylindrical coordinate system centered on a laser-drilled hole; at a specific location in the epithelial plane, the azimuthal direction is in the epithelial plane, but perpendicular to the radial direction (which is parallel to a line from the location to the hole)

*creep compliance* – function that characterizes the time-dependent response of a material to a unit step change in stress *laser hole-drilling* – process in which a tightly focused, high peak-power laser pulse is targeted onto a surface to locally destroy the surface's mechanical integrity; typically occurs via local plasma generation, subsequent confined boiling, bubble expansion and collapse

*relaxed strain* – the change in strain after mechanical modification of a surface; typically after cutting or hole-drilling

*relaxed displacement* – field of displacements that accompanies relaxed strain

## Alphabetized Table of Symbols

$D$ = displacement coefficient or prefactor in power-law fits of recoil kinetics
$d_0$ = cell edge position at the time of ablation
$d_1, d_2$ = displacement coefficients in double-exponential fits of recoil kinetics
$E$ = Young's modulus of the epithelium
$G_0$ = scale factor for stiffness in a power-law creep compliance
$m$ = cell edge velocity before ablation
$r, \theta$ = radial and azimuthal coordinates in a cylindrical coordinate system centered on the hole
$r_0$ = initial distance of a tracked cell edge from the laser-drilled hole
$R_0$ = initial size of the laser-drilled hole
$R_{xx}$ = ratio of parameter xx in Stage 14 to that in Stage 13; xx may be $\sigma C$ for cell-center stress, $\sigma E$ for cell-edge stress, G' for tissue stiffness or $\eta$ for viscous damping coefficient
$t_0$ = time of ablation
$t_1, t_2$ = time constants in double-exponential fits of recoil kinetics
$u_r, u_\theta$ = post-ablation relaxed displacements in the radial and azimuthal directions
$u_{tr}, \theta_{tr}$ = magnitude and direction of post-ablation rigid-body translation
$v_0$ = initial post-ablation recoil velocity
$\langle v_{0-2s} \rangle$ = average recoil velocity from zero to two seconds after ablation
$x(t)$ = time-dependent position of a cell edge
$\alpha$ = exponent in power-law fits of recoil kinetics
$\gamma_0$ = initial post-ablation strain rate
$\Delta$ = strain coefficient or prefactor for power-law fits of recoil kinetics
$\Delta x(t)$ = time-dependent displacement of a cell edge
$\Delta \varepsilon_r, \Delta \varepsilon_\theta$ = post-ablation relaxed strains in the radial and azimuthal directions
$\varepsilon(t)$ = time-dependent strain in a creep experiment
$\eta$ = effective viscous drag coefficient
$\nu$ = Poisson ratio of the epithelium
$\rho(\theta)$ = normalized number density of cell edges parallel to the direction q
$\sigma$ = applied stress in a creep experiment
$\sigma_x, \sigma_y$ = pre-ablation biaxial stresses in the epithelium
$\tau_0$ = scale factor for time in a power-law creep compliance

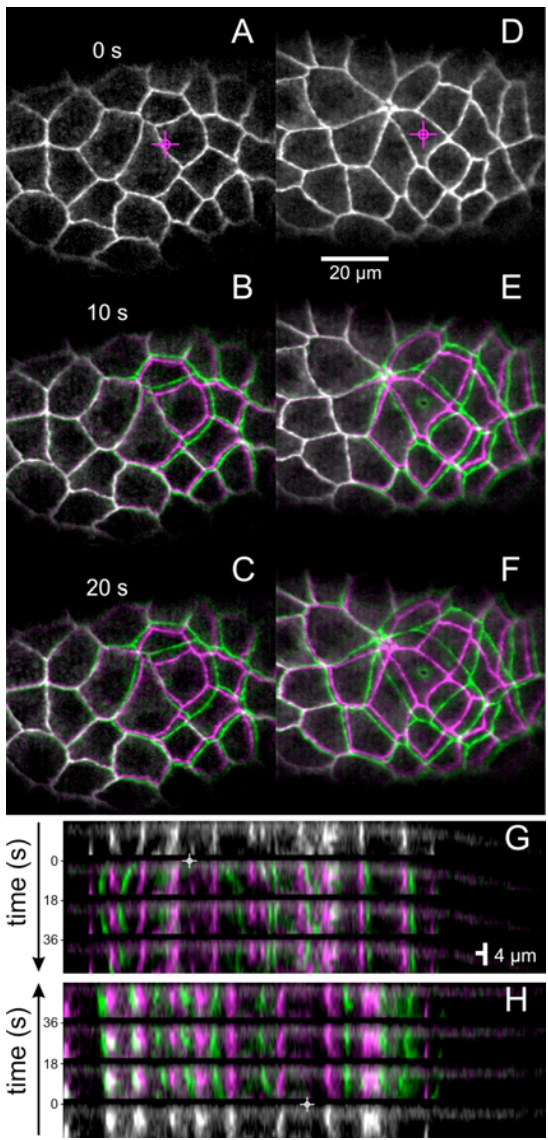

**Figure 1.** Mechanical response of the amnioserosa to single-pulse laser ablation. **(A-C)** are confocal images taken before, 10 s and 20 s after ablation. The post-ablation images are overlays comparing the cell border positions before **(magenta)** and after ablation **(green)**. The laser was targeted to the cell edge under the **crosshairs** in A. **(D-F)** are a similar series in which the laser targets a single cell's apical surface. For each image, anterior is to the right. **(G-H)** are time-lapse cross-sectional images through the amnioserosa (dorsal up). Cell borders appear as nearly vertical bright lines. The **crosshairs** mark the time and location of ablation. The cross-sections in **(G)** were collected dorsal-to-ventral, so time increases as one goes down a single image. The cross-sections in **(H)** were collected ventral-to-dorsal, so time increases as one goes up a single image. Movies S1 and S2 corresponding to (A-C) and (D-F) are available as supplemental material.

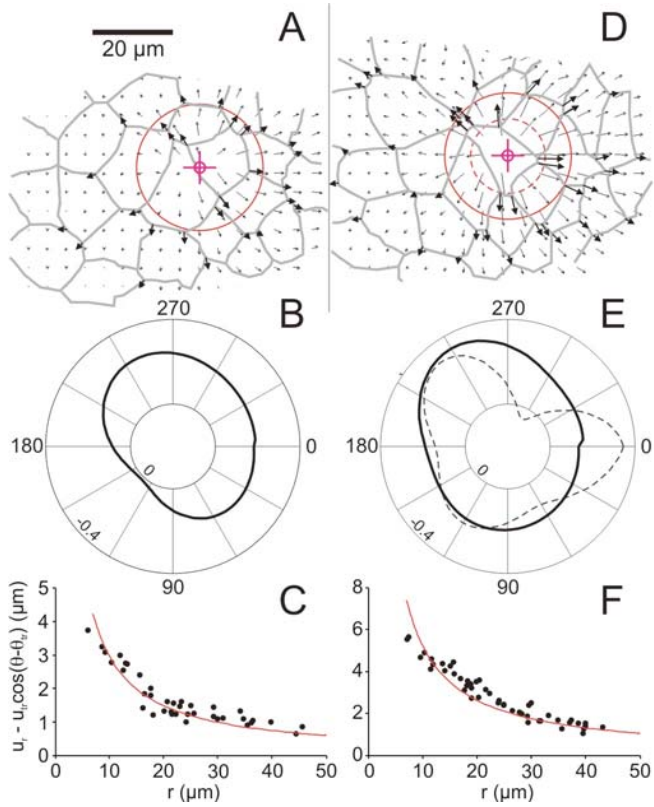

**Figure 2.** Displacement and relaxed strain patterns at 20 s after ablation. **(A,D)** are displacement vector fields from B-spline warps **(light arrows)** or from tracking specific cellular triple junctions **(bold arrows)**. The ablated spot is denoted by **(crosshairs)** with concentric rings at distances of 9.4 µm **(dashed)** and 15.6 µm **(solid)**. The scale bar in A applies to both panels. **(B,E)** are corresponding polar plots of the relaxed strains for $r$ = 9.4 µm **(dashed)** and 15.6 µm **(solid)**. The relaxed strain scale ranges from +0.2 at the center of the plots to -0.4 at the outer ring. **(C,F)** are plots of the corrected radial displacements. Each point corresponds to a triple-junction displacement and the line is the best fit to the isotropic version of Eq 5. Note that **(A-C)** correspond to Fig 1A-C and **(D-F)** correspond to Fig 1D-F.

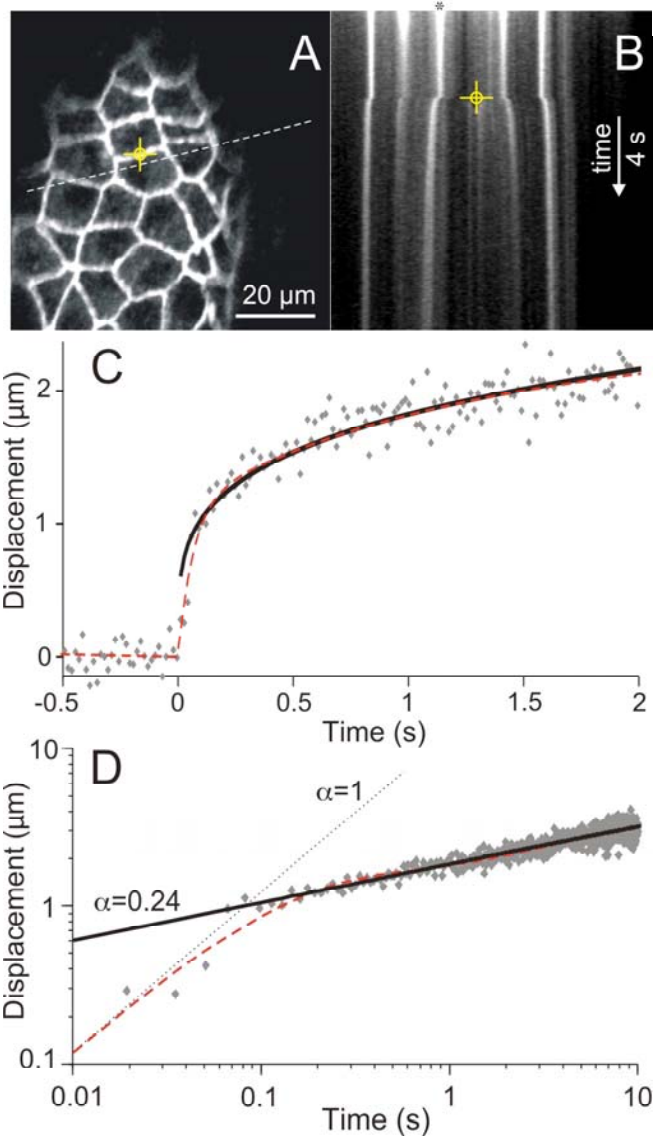

**Figure 3.** Line-scan kymographs for measuring the fast recoil after laser ablation. (**A**) is a confocal fluorescent image of amnioserosa cells before ablation. The targeted cell edge is marked with **crosshairs** and the **dashed line** marks the line that will be repeatedly scanned. The repeated line scans are used to build the kymograph (**B**). The horizontal axis is positioned along the marked line; the vertical axis is time. The cell borders crossed by the line scan appear as bright bands that are nearly vertical before ablation. After ablation (marked in time and space by the **crosshairs**), these bands fan out as the cells recoil. The cell border marked with an **asterisk** was tracked to yield the displacement versus time plot (**C**), shown on a log-log scale in (**D**). The cell edge displacement was fit with a piecewise continuous function: linear before ablation and either a double exponential (**dashed line**) or power-law (**solid line**) after ablation.

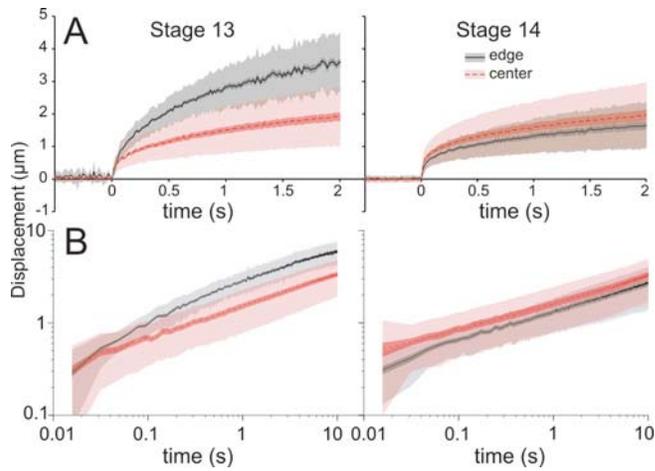

**Figure 4**. Dependence of the recoil dynamics on wound site and developmental stage. Left panels are from stage 13 embryos; right panels are from stage 14. **(A)** Average dynamic recoils for cell-edge wounds **(grey-solid)** or cell-center wounds **(red-dashed)**. The lightly shaded regions represent the standard deviations; the darker regions represent the standard errors of the means. **(B)** Same, but on a log-log scale.

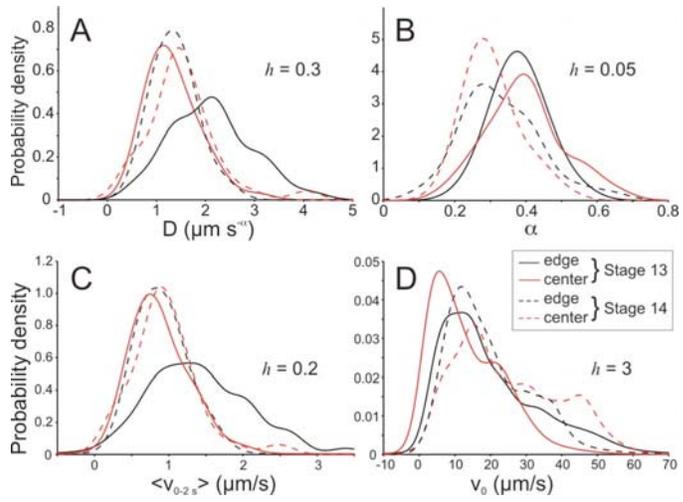

**Figure 5.** Kernel density estimates for the recoil parameter distributions. Each panel has four distributions (two wound sites by two developmental stages) for a single parameter. For each parameter, the kernel width (*h*) is noted.

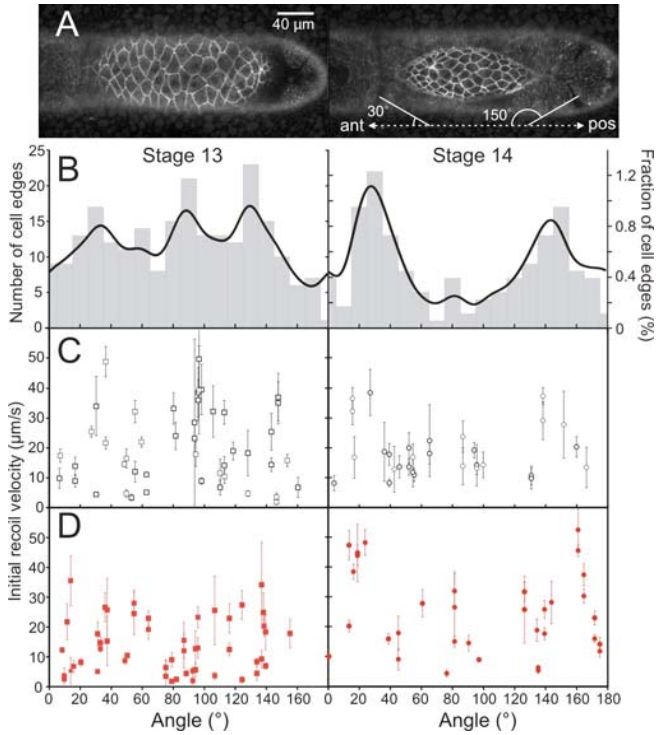

**Figure 6**. Angular dependence of the dynamic recoils. The left panels are from stage 13 embryos; the right ones are from stage 14. **(A)** Confocal fluorescent images of amnioserosa cells (anterior to the left). **(B)** Histograms and kernel density estimates for the stage-dependent angular distribution of cell edges. Angles are defined as in A. **(C)** The initial recoil velocity is plotted versus the tracked recoil direction for cell-edge ablation. The tracked direction was always parallel to the ablated edge. **(D)** Similar plot for cell-center wounds. Each point represents a single experiment with error bars at 95% confidence limits.

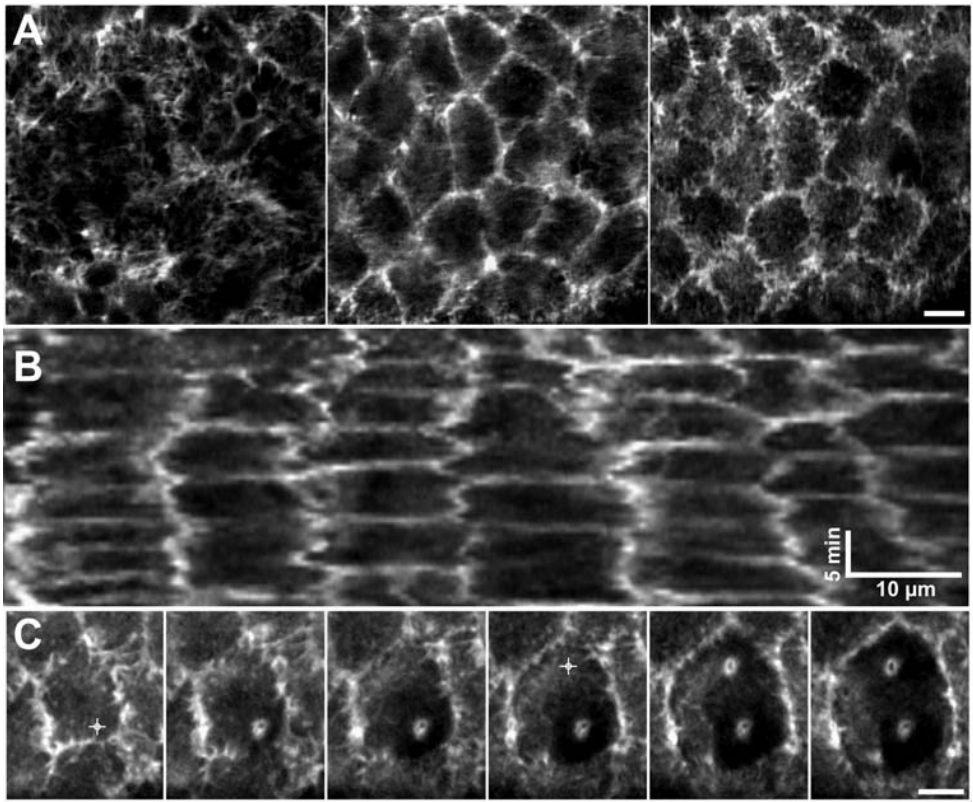

**Figure 7.** The apical actin network in GFP-moesin embryos. **(A)** Confocal fluorescent images of amnioserosa cells taken in different imaging planes: basal, middle and apical (from left to right). **(B)** Long-time kymograph of contractile waves coupled to actin accumulations in the amnioserosa (time on the vertical axis). **(C)** Time-series of apical-plane images during a double-wounding experiment. Ablation targets the crosshairs just after the 1$^{st}$ and 4$^{th}$ images (12 s between images). Images are also available as supplemental Movie S7. Each horizontal scale bar is 10 µm.

SUPPLEMENTARY INFORMATION

| Distance above adherens junctions | Basal surface cut | Recoil observed |
|---|---|---|
| 4 μm | 4/5 | 5/5 |
| 5 μm | 1/3 | 3/3 |
| 6 μm | 1/3 | 3/3 |
| 7 μm | 0/2 | 1/2 |
| 8 μm | 0/3 | 1/3 |

**Table S1.** Results from ablation experiments in which the laser was focused a given distance above the cells' adherens junctions. For each distance, the table compiles the fraction of experiments in which the basal surface was cut and/or recoil was observed.

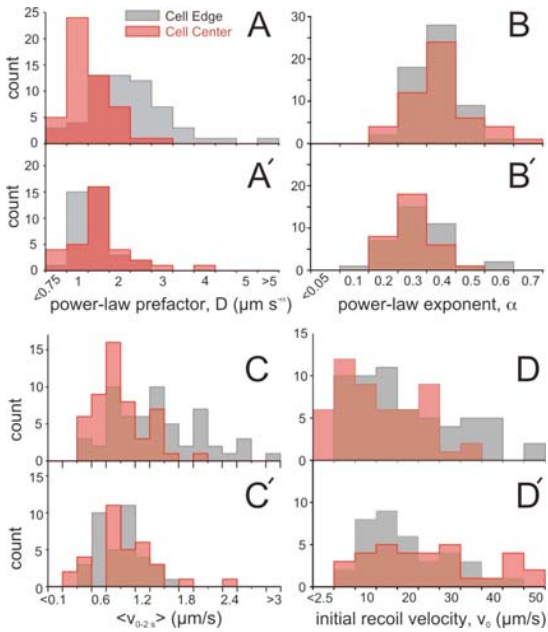

**Figure S1.** Histograms of the recoil parameters. Each panel has two distributions (cell-edge and cell-center wounds) for a single parameter. The unprimed panels correspond to stage 13, primed to stage 14.

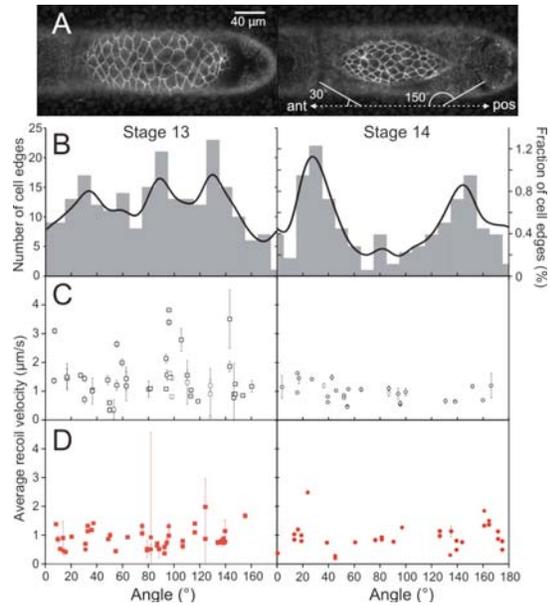

**Figure S2.** Angular dependence of the recoil dynamics as measured by the average recoil velocity during the first two seconds after ablation $\langle v_{0\text{-}2s}\rangle$. The left panels are from stage 13 embryos; the right ones are from stage 14. **(A)** Confocal fluorescent images of amnioserosa cells (anterior to the left). **(B)** Histograms and kernel density estimates (**solid lines**) for the stage-dependent angular distribution of cell edges. Angles are defined as in A. **(C)** $\langle v_{0\text{-}2s}\rangle$ is plotted versus the tracked recoil direction for cell-edge ablation. The tracked direction was always parallel to the ablated edge. **(D)** Similar plot of $\langle v_{0\text{-}2s}\rangle$ for cell-center wounds. Each point represents a single experiment with error bars at 95% confidence limits.

Supplemental Note 1 – Coupled constraints on mechanical properties during dorsal closure

If one assumes that the mechanical properties of the epithelium ($G_0$, $t_0$, $\alpha$ and $\eta$) are independent of wound site, then the recoil parameters enable two estimates for the ratio of cell-edge stress to cell-center stress in each developmental stage. For example, in stage 13 (similarly for stage 14),

$$R_{\sigma 13} = \frac{\sigma_{E13}}{\sigma_{C13}} = \frac{\Delta_{E13}}{\Delta_{C13}} \quad \text{or} \quad \frac{\gamma_{0,E13}}{\gamma_{0,C13}} \qquad \text{Eqn S1.}$$

The two estimates of each ratio are quite close to one another, so we take the averages: $R_{\sigma 13} = 1.53 \pm 0.11$ and $R_{\sigma 14} = 0.74 \pm 0.08$. These two ratios can be combined into a grand ratio that summarizes the coupled constraint on how the cell-edge and cell-center stresses may change:

$$\frac{R_{\sigma 13}}{R_{\sigma 14}} = \frac{\sigma_{E13}/\sigma_{C13}}{\sigma_{E14}/\sigma_{C14}} = \frac{\sigma_{C14}/\sigma_{C13}}{\sigma_{E14}/\sigma_{E13}} = \frac{R_{\sigma C}}{R_{\sigma E}} = \Gamma = 2.06 \pm 0.28 \qquad \text{Eqn S2.}$$

One can then relate the constraints on the stresses and those on the mechanical parameters

$$R_{\Delta E} R_{\Delta C} = \frac{\Delta_{E14}}{\Delta_{E13}} \frac{\Delta_{C14}}{\Delta_{C13}} = \frac{R_{\sigma E}}{R_{G'}} \frac{R_{\sigma C}}{R_{G'}} = \frac{\Gamma R_{\sigma E}^2}{R_{G'}^2} \qquad \text{Eqn S3,}$$

$$R_{\gamma E} R_{\gamma C} = \frac{\gamma_{0,E14}}{\gamma_{0,E13}} \frac{\gamma_{0,C14}}{\gamma_{0,C13}} = \frac{R_{\sigma E}}{R_{\eta}} \frac{R_{\sigma C}}{R_{\eta}} = \frac{\Gamma R_{\sigma E}^2}{R_{\eta}^2} \qquad \text{Eqn S4,}$$

where $R_{G'} = (G_0 t_0^\alpha)_{14} / (G_0 t_0^\alpha)_{13}$ and $R_{\eta} = \eta_{14}/\eta_{13}$. Rearranging Eqn S3 and S4 yields:

$$\frac{R_{G'}}{R_{\sigma E}} = \left(\frac{\Gamma}{R_{\Delta E} R_{\Delta C}}\right)^{1/2} = 1.24 \pm 0.08 \qquad \text{Eqn S5,}$$

$$\frac{R_{\eta}}{R_{\sigma E}} = \left(\frac{\Gamma}{R_{\gamma E} R_{\gamma C}}\right)^{1/2} = 0.77 \pm 0.09 \qquad \text{Eqn S6.}$$

These constraints are listed in Table II of the main text.